\documentclass[a4paper,10pt]{article}
\pdfoutput=1
\usepackage{float}
\usepackage{longtable}
\usepackage{titling}

\usepackage[utf8]{inputenc} 
\usepackage[T1]{fontenc} 
\usepackage[british]{babel} 
\usepackage{subfigure}
\usepackage[final]{pdfpages} 

\usepackage[colorlinks=true,urlcolor=black,linkcolor=black]{hyperref} 

\usepackage{lmodern} 
\usepackage[paper=a4paper]{geometry} 

\usepackage{amsmath} 
\usepackage{esint} 
\usepackage{amssymb} 
\usepackage{esvect} 
\usepackage{tensor}
\usepackage{mathrsfs} 

\usepackage{array}
\usepackage{cellspace} 
\usepackage{makecell}
\usepackage{multirow}

\usepackage{tikz}
\usetikzlibrary{patterns}
\usetikzlibrary{calc,arrows,positioning}
\usetikzlibrary{arrows,shadows}
\usetikzlibrary{decorations.pathmorphing}	
\usetikzlibrary{decorations.markings}

\usepackage{textcomp}
\usepackage{graphicx}
\usepackage{url}
\usepackage{appendix}
\usepackage{subfigure}
\usepackage{pgfplots}
\usepackage{verbatim} 

\usepgfplotslibrary{patchplots}
\pgfplotsset{compat=1.3}

\usepackage{epstopdf}
\usepackage{listings}
\usepackage{lipsum}
\usepackage{colortbl}
\usepackage{qtree}
\usepackage{color}
\usepackage{gensymb}
\usepackage{enumerate}
\usepackage{numprint}
\usepackage{caption}
\usepackage[nottoc, notlof, notlot]{tocbibind} 
\usepackage{rotating} 

\usepackage{hhline}

\usepackage{vmargin}
\usepackage[final]{pdfpages}
\usepackage{titlesec}
\usepackage{stmaryrd}
\usepackage{delarray}

\newcommand{\dif}{\mathrm{d}}

\newcommand{\pa}{\partial}
\renewcommand{\ge}{\geqslant}
\renewcommand{\le}{\leqslant}

\title{\textbf{Tangent method for the arctic curve arising from freezing boundaries}}
\author{{Bryan Debin and Philippe Ruelle}\\
\\
\textit{Universit\'e catholique de Louvain}\\
\textit{Institut de Recherche en Math\'ematique et Physique}\\
\textit{Chemin du Cyclotron 2, 1348 Louvain-la-Neuve, Belgium}\\
\\
{\tt bryan.debin\,@\,uclouvain.be,}
{\tt philippe.ruelle\,@\,uclouvain.be}
}
\date{\today}

\begin{document}

\maketitle

\begin{abstract}
In the paper \cite{DiFrancesco2018ArbitraryStartingPoint}, the authors study the arctic curve arising in random tilings of some planar domains with an arbitrary distribution of defects on one edge. Using the tangent method they derive a parametric equation for portions of arctic curve in terms of an arbitrary piecewise differentiable function that describes the defect distribution. When this distribution presents "freezing" intervals, other portions of arctic curve appear and typically have a cusp. These freezing boundaries can be of two types, respectively with maximal or minimal density of defects. Our purpose here is to extend the tangent method derivation of \cite{DiFrancesco2018ArbitraryStartingPoint} to include these portions, hence answering the open question stated in \cite{DiFrancesco2018ArbitraryStartingPoint}.
\end{abstract}

\section{Introduction}

Spatial phase separation phenomena have attracted a lot of attention recently, among which those appearing in random tiling problems. One of the first examples concerns the domino (or dimer) tiling of aztec diamonds
\cite{Jockusch1998random}, where a liquid, disordered region and a solid, ordered region are separated by an interface that becomes infinitely sharp in the scaling limit, giving rise to the famous arctic circle. More instances of this phenomenon occur in other statistical physics problems, such as the six-vertex model with domain-wall boundary conditions \cite{Colomo2016TangentMethod}, the imaginary time evolution of some one-dimensional quantum spin chains from a domain-wall initial state \cite{Allegra2016inhomogeneous}, or the two-periodic tilings of the aztec diamond \cite{DK17}. Here we will exclusively focus on the tiling model considered in \cite{DiFrancesco2018ArbitraryStartingPoint}. Rather than a single model, it is in fact a rich family of models, as its scaling limit is specified in terms of a real function defined on an interval $[0,1]$.

The shape of the arctic curve was already determined rigorously in \cite{Petrov2014} for a particular defect distribution (freezing intervals only) and then in \cite{Duse2015} for a general distribution. Our main objective is to complete the derivation of the arctic curve made in \cite{DiFrancesco2018ArbitraryStartingPoint} by extending their argument based on the recently introduced tangent method \cite{Colomo2016TangentMethod}. The emphasis is therefore not on the shape of the arctic curve itself but rather on the full validation of the tangent method in these somewhat peculiar cases where pieces of arctic curves show a cusp.

This note refers directly to the material presented in \cite{DiFrancesco2018ArbitraryStartingPoint}, and for convenience, we will use the same notations wherever possible. In order to make the article minimally self-contained, we quickly review the model under consideration and the tangent method, respectively in the second and third sections. Section 2 also contains the statement of the problem left open in \cite{DiFrancesco2018ArbitraryStartingPoint}. Section 4 contains a proof of this open problem regarding the tangent method based derivation of the arctic curve forming above two types of freezing boundaries. Our derivation turns out to be rather short, and may be applied uniformly to all portions of the arctic curve.

\section{Definition of the model}

The model examined in \cite{DiFrancesco2018ArbitraryStartingPoint} is primarily defined in terms of $n$ non-intersecting lattice paths (NILP) contained in a quadrant. The $n$ paths are labelled by an integer $i$ between 1 and $n$. The $i$-th path starts from the integer position $O_i=(a_i,0)$ on the positive horizontal axis, and ends on the vertical axis at position $E_i=(0,i)$. An extra path, reduced to a single site, is conventionally added at the origin ($a_0=0$).

The paths, between their starting and ending points, can only make two kinds of unit steps, either north or west (steps $(0,1)$ or $(-1,0)$), and are constrained to be non-intersecting. This last condition implies that the arbitrary but fixed integers $(a_i)_i$ form a strictly increasing sequence, $0 < a_1 < a_2 < \ldots < a_n$. The collection of $n$ paths is therefore entirely contained in the rectangle of size $a_n \times n$. We are interested in NILP chosen at random among all the allowed configurations. Since the number of unit steps of each type is entirely determined by the set of $a_i$'s, we give a uniform weight to the allowed configurations of  NILP, without loss of generality.

Each NILP configuration can be associated with a specific tiling of a domain that is close to the rectangle mentioned above (see Figure \ref{fig_definition_model}). The tiles are of three types, called U, R and F, as they correspond to the Upper, Right and Front faces of a cuboid drawn in perspective. The correspondence between NILP and the tilings is shown on Figure \ref{fig_definition_model}, with the U-tiles carrying the west steps and the R-tiles carrying the north steps,  while the F-tiles carry no path segment (see  inset $\mathbf{A}$ of this figure). 

\newcommand{\Ff}{--++(1,0)--++(0,1)--++(-1,0)--++(0,-1);}
\newcommand{\Rr}{--++(1,1)--++(0,1)--++(-1,-1)--++(0,-1);}
\newcommand{\Uu}{--++(1,0)--++(1,1)--++(-1,0)--++(-1,-1);}

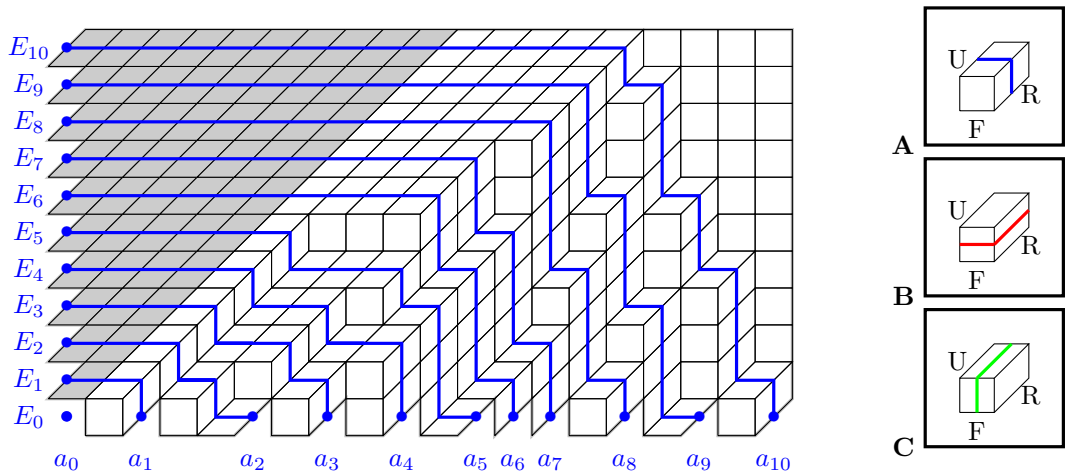
\begin{figure}[h!]
\begin{center}
\begin{tikzpicture}[scale=0.7]
\begin{scope}[xshift=9cm,yshift=1cm,scale=0.65]
\draw[very thick] (18-7,7-1+0.4) node[left]{$\mathbf{A}$}rectangle++(4,4);
\draw[very thick] (18-7,7-1-4) node[left]{$\mathbf{B}$}rectangle++(4,4);
\draw[very thick] (18-7,7-1-8-0.4) node[left]{$\mathbf{C}$}rectangle++(4,4);
\draw (12,7+0.4) --++(0.5,0)node[below]{F}--++(0.5,0)--++(0,1)--++(-1,0)--++(0,-1);
\draw (12,8+0.4)--++(0.5,0.5)node[left]{U}--++(0.5,0.5)--++(1,0)--++(-1,-1);
\draw (13,7+0.4)--++(0.5,0.5)node[right]{R}--++(0.5,0.5)--++(0,1);
\draw[very thick,blue] (12+0.5,7+1+0.5+0.4)--++(1,0);
\draw[very thick,blue] (12+1.5,7+1+0.5+0.4)--++(0,-1);
\draw (12,7-4) --++(0.5,0)node[below]{F}--++(0.5,0)--++(0,1)--++(-1,0)--++(0,-1);
\draw (12,8-4)--++(0.5,0.5)node[left]{U}--++(0.5,0.5)--++(1,0)--++(-1,-1);
\draw (13,7-4)--++(0.5,0.5)node[right]{R}--++(0.5,0.5)--++(0,1);
\draw[very thick,red] (12,7.5-4)--++(1,0);
\draw[very thick,red] (13,7.5-4)--++(1,1);
\draw (12,7-8-0.4) --++(0.5,0)node[below]{F}--++(0.5,0)--++(0,1)--++(-1,0)--++(0,-1);
\draw (12,8-8-0.4)--++(0.5,0.5)node[left]{U}--++(0.5,0.5)--++(1,0)--++(-1,-1);
\draw (13,7-8-0.4)--++(0.5,0.5)node[right]{R}--++(0.5,0.5)--++(0,1);
\draw[very thick,green] (12.5,7-8-0.4)--++(0,1);
\draw[very thick,green] (12.5,8-8-0.4)--++(1,1);
\end{scope}

\begin{scope}[scale=0.7]
\begin{scope}[xshift=-0.5cm,yshift=-0.5cm]
\draw[very thick, gray!50] (1,0)--++(1,0)--++(1,1)--++(0,-1)--++(2,0)--++(1,1)--++(0,-1)--++(0.5,0) --++(0.5,0)--++(1,1)--++(0,-1)--++(1,0)--++(1,1)--++(0,-1)--++(1,0)--++(1,1)--++(0,-1)--++(1,1)--++(0,-1)--++(1,1)--++(0,-1)--++(1,0)--++(1,1)--++(0,-1)--++(1,0)--++(1,1)--++(0,-1) --++(0.5,0) --++(0.5,0)--++(1,1)--++(0,4)--++(0,0.5)--++(0,0.5)--++(0,5)--++(-19,0)--++(-1,-1)--++(1,0)--++(-1,-1)--++(1,0)--++(-1,-1)--++(1,0)--++(-1,-1)--++(1,0)--++(-1,-1)--++(1,0)--++(-0.5,-0.5)--++(-0.5,-0.5)--++(1,0)--++(-1,-1)--++(1,0)--++(-1,-1)--++(1,0)--++(-1,-1)--++(1,0)--++(-1,-1)--++(1,0)--++(0,-1);
\end{scope}

\begin{scope}[yshift=-0.5cm, xshift=-0.5cm]

\draw[fill=gray!40] (0,1) \Uu \draw[fill=gray!40] (0,2) \Uu \draw[fill=gray!40] (0,3) \Uu \draw[fill=gray!40] (0,4) \Uu\draw[fill=gray!40] (0,5) \Uu \draw[fill=gray!40] (0,6) \Uu \draw[fill=gray!40] (0,7) \Uu \draw[fill=gray!40] (0,8) \Uu \draw[fill=gray!40] (0,9) \Uu \draw[fill=gray!40] (0,10) \Uu 

                          \Uu \draw[fill=gray!40] (1,2) \Uu \draw[fill=gray!40] (1,3) \Uu \draw[fill=gray!40] (1,4) \Uu\draw[fill=gray!40] (1,5) \Uu \draw[fill=gray!40] (1,6) \Uu \draw[fill=gray!40] (1,7) \Uu \draw[fill=gray!40] (1,8) \Uu \draw[fill=gray!40] (1,9) \Uu \draw[fill=gray!40] (1,10) \Uu 
                          
\draw[fill=gray!40] (2,3) \Uu \draw[fill=gray!40] (2,4) \Uu\draw[fill=gray!40] (2,5) \Uu \draw[fill=gray!40] (2,6) \Uu \draw[fill=gray!40] (2,7) \Uu \draw[fill=gray!40] (2,8) \Uu \draw[fill=gray!40] (2,9) \Uu \draw[fill=gray!40] (2,10) \Uu 

                              \draw[fill=gray!40] (3,4) \Uu\draw[fill=gray!40] (3,5) \Uu \draw[fill=gray!40] (3,6) \Uu \draw[fill=gray!40] (3,7) \Uu \draw[fill=gray!40] (3,8) \Uu \draw[fill=gray!40] (3,9) \Uu \draw[fill=gray!40] (3,10) \Uu 
                              
\draw[fill=gray!40] (4,5) \Uu \draw[fill=gray!40] (4,6) \Uu \draw[fill=gray!40] (4,7) \Uu \draw[fill=gray!40] (4,8) \Uu \draw[fill=gray!40] (4,9) \Uu \draw[fill=gray!40] (4,10) \Uu 

                              \draw[fill=gray!40] (5,6) \Uu \draw[fill=gray!40] (5,7) \Uu \draw[fill=gray!40] (5,8) \Uu \draw[fill=gray!40] (5,9) \Uu \draw[fill=gray!40] (5,10) \Uu 
                               
\draw[fill=gray!40] (6,7) \Uu \draw[fill=gray!40] (6,8) \Uu \draw[fill=gray!40] (6,9) \Uu \draw[fill=gray!40] (6,10) \Uu 

                              \draw[fill=gray!40] (7,8) \Uu \draw[fill=gray!40] (7,9) \Uu \draw[fill=gray!40] (7,10) \Uu 
                              
\draw[fill=gray!40] (8,9) \Uu \draw[fill=gray!40] (8,10) \Uu 

\draw[fill=gray!40] (9,10) \Uu

\draw (1,0) \Ff \draw (2,0) \Rr \draw (3,0) \Ff \draw (4,0) \Rr \draw (4,0) \Uu \draw (6,0) \Ff \draw (7,0) \Rr \draw (8,0) \Ff \draw (9,0) \Rr \draw (10,0) \Uu \draw (10,0) \Rr \draw (12,0) \Rr \draw (13,0) \Rr \draw (14,0) \Ff \draw (15,0) \Rr \draw (16,0) \Uu \draw (16,0) \Rr \draw (18,0) \Ff \draw (19,0) \Rr

\draw (1,1) \Uu \draw (3,1) \Rr \draw (3,1) \Uu \draw (5,1) \Ff \draw (6,1) \Rr \draw (6,1) \Uu \draw (8,1) \Ff \draw (9,1) \Rr \draw (10,1) \Rr \draw (11,1) \Rr \draw (11,1) \Uu \draw (13,1) \Rr \draw (14, 1) \Uu \draw (14,1) \Rr \draw (16,1) \Rr \draw (17,1) \Ff \draw (18,1) \Uu \draw (18,1) \Rr

\draw (2,2) \Uu \draw (4,2) \Uu \draw (4,2) \Rr \draw (5,2) \Uu \draw (7,2) \Uu \draw (7,2) \Rr \draw (8,2) \Uu \draw (10,2) \Rr \draw (11,2) \Rr \draw (12,2) \Rr \draw (12,2) \Uu \draw (14,2) \Rr \draw (15,2) \Ff \draw (16,2) \Rr \draw (17,2) \Ff \draw (18,2) \Rr \draw (19,2) \Ff

\draw (3,3) \Uu \draw (5,3) \Uu \draw (5,3) \Rr \draw (6,3) \Uu  \draw (8,3) \Ff \draw (9,3) \Uu \draw (9,3) \Rr \draw (11,3) \Rr \draw (12,3) \Rr \draw (13,3) \Ff \draw (14,3) \Rr \draw (15,3) \Rr \draw (15,3) \Uu \draw (17,3) \Ff \draw (18,3) \Rr \draw (19,3) \Ff

\draw (4,4) \Uu \draw (6,4) \Uu \draw (6,4) \Rr \draw (7,4) \Uu \draw (8,4) \Uu \draw (10,4) \Uu \draw (10,4) \Rr \draw (12,4) \Rr \draw (13,4) \Rr \draw (13,4) \Uu \draw (15,4) \Rr \draw (16,4) \Ff \draw (17,4) \Uu \draw (17,4) \Rr \draw (19,4) \Ff

\draw (5,5) \Uu \draw (7,5) \Ff \draw (8,5) \Ff \draw (9,5) \Ff \draw (10,5) \Rr \draw (11,5) \Rr \draw (11,5) \Uu \draw (13,5) \Rr \draw (14,5) \Ff \draw (15,5) \Rr \draw (16,5) \Ff \draw (17,5) \Rr \draw (18,5) \Ff \draw (19,5) \Ff

\draw (6,6) \Uu \draw (7,6) \Uu \draw (8,6) \Uu \draw (9,6) \Uu \draw (11,6) \Rr \draw (12,6) \Ff \draw (13,6) \Rr \draw (14,6) \Uu \draw (14,6) \Rr \draw (16,6) \Uu \draw (16,6) \Rr \draw (18,6) \Ff \draw (19,6) \Ff

\draw (7,7) \Uu \draw (8,7) \Uu \draw (9,7) \Uu \draw (10,7) \Uu \draw (12,7) \Ff \draw (13,7) \Rr \draw (14,7) \Rr \draw (15,7) \Ff \draw (16,7) \Rr \draw (17,7) \Ff \draw (18,7) \Ff \draw (19,7) \Ff 

\draw (8,8) \Uu \draw (9,8) \Uu \draw (10,8) \Uu \draw (11,8) \Uu \draw (12,8) \Uu \draw (14,8) \Rr \draw (15,8) \Ff \draw (16,8) \Rr \draw (17,8) \Ff \draw (18,8) \Ff \draw (19,8) \Ff

\draw (9,9) \Uu \draw (10,9) \Uu \draw (11, 9) \Uu \draw (12,9) \Uu \draw (13,9) \Uu \draw (15,9) \Uu \draw (15,9) \Rr \draw (17,9) \Ff \draw (18,9) \Ff \draw (19,9) \Ff

\draw (10,10) \Uu \draw (11,10) \Uu \draw (12,10) \Uu \draw (13,10) \Uu \draw (14,10) \Uu \draw (16,10) \Ff \draw (17,10) \Ff \draw (18,10) \Ff \draw (19,10) \Ff
\end{scope}

\draw[blue] (0,0) node{$\bullet$} node[yshift=-0.6cm]{$a_0$};
\draw[very thick,blue] (2,0) node{$\bullet$} node[yshift=-0.6cm]{$a_1$}--++(0,1)--++(-2,0);
\draw[very thick,blue] (5,0) node{$\bullet$} node[yshift=-0.6cm]{$a_2$}--++(-1,0)--++(0,1)--++(-1,0)--++(1,0)--++(-1,0)--++(0,1)--++(-3,0);
\draw[very thick,blue] (7,0) node{$\bullet$} node[yshift=-0.6cm]{$a_3$}--++(0,1)--++(-1,0)--++(0,1)--++(-2,0)--++(1,0)--++(-1,0)--++(0,1)--++(-4,0);
\draw[very thick,blue] (9,0) node{$\bullet$} node[yshift=-0.6cm]{$a_4$}--++(0,1)--++(0,1)--++(-2,0)--++(1,0)--++(-1,0)--++(0,1)--++(-2,0)--++(0,1)--++(-5,0);
\draw[very thick,blue] (11,0) node{$\bullet$} node[yshift=-0.6cm]{$a_5$}--++(-1,0)--++(0,3)--++(-1,0)--++(0,1)--++(-3,0)--++(0,1)--++(-6,0);
\draw[very thick,blue] (12,0) node{$\bullet$} node[yshift=-0.6cm]{$a_6$}--++(0,1)--++(-1,0)--++(0,3)--++(-1,0)--++(0,2)--++(-10,0);
\draw[very thick,blue] (13,0) node{$\bullet$} node[yshift=-0.6cm]{$a_7$}--++(0,2)--++(-1,0)--++(0,3)--++(-1,0)--++(0,2)--++(-11,0);
\draw[very thick,blue] (15,0) node{$\bullet$} node[yshift=-0.6cm]{$a_8$}--++(0,1)--++(-1,0)--++(0,3)--++(-1,0)--++(0,2)--++(0,2)--++(-13,0);
\draw[very thick,blue] (17,0) node{$\bullet$} node[yshift=-0.6cm]{$a_9$}--++(-1,0)--++(0,3)--++(-1,0)--++(0,3)--++(-1,0)--++(0,3)--++(-14,0);
\draw[very thick,blue] (19,0) node{$\bullet$} node[yshift=-0.6cm]{$a_{10}$}--++(0,1)--++(-1,0)--++(0,3)--++(-1,0)--++(0,2)--++(-1,0)--++(0,3)--++(-1,0)--++(0,1)--++(-15,0);
\foreach \x in {0,1,2,...,10} {
\draw[blue] (0,\x) node{$\bullet$} node[xshift=-0.5cm]{$E_{\x}$};}
\end{scope}

\end{tikzpicture}
\end{center}
\caption{Particular configuration of NILP for the model under consideration and the associated tiling on a roughly rectangular domain. The tiles drawn in grey are deterministically frozen. The insets $\mathbf{A}$, $\mathbf{B}$ and $\mathbf{C}$ describe the rule to translate a tiling in terms of NILP  respectively of type 1, 2 and 3. We will call "defects" the triangular inclusions on the lower boundary corresponding to the starting points $O_i$ with $i=1,\cdots, n$.}
\label{fig_definition_model}
\end{figure}

The paths discussed so far will be called "of type 1". Indeed the tilings with which they are associated can as well be described by two other types of paths, shown on Figure \ref{fig_definition_model} in red (type 2) and in green (type 3). An essential feature of this triple path description is that a region covered with one type of tile is associated with a total abscence of path of one type, something that is crucial to apply the tangent method. In addition to be characterized by different steps, the three types of paths also have different starting and ending points, see \cite{DiFrancesco2018ArbitraryStartingPoint} and below for more details.

Due to the tightly packed arrangement of ending points $E_i$, there is a region that is deterministically tiled  by a regular (frozen) pattern of U-tiles. More interestingly, there are also statistically frozen regions that appear in the vast majority of the tilings.  As already suggested by Figure \ref{fig_definition_model}, two frozen regions tend to form in the upper left and upper right corners of the rectangle, each being tiled by only one type of tile, U or F. For a generic distribution of starting points $a_i$, no other frozen region appears. In the scaling limit, an arctic curve separates these two frozen regions from the entropic region. The precise shape of this arctic curve has been completely and explicitly determined in \cite{DiFrancesco2018ArbitraryStartingPoint}; it only depends on the function $\alpha(u)$ that characterizes, in the scaling limit, the distribution of the points $a_i$.

\begin{figure}[h!]
\includegraphics[scale=1.45]{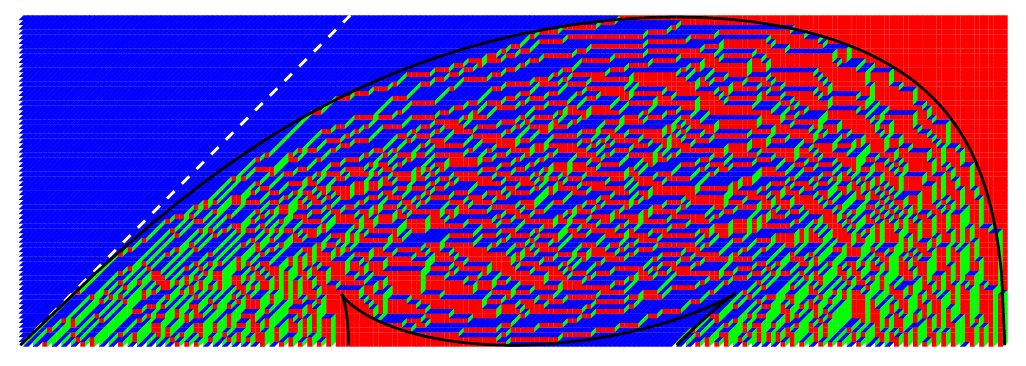}
\caption{Random tiling of a domain with a macroscopic gap in the defect distribution of the lower edge. The intervals on the left and on the right of the gap have the same length and contain one defect every second site. The number of paths (defects) is $n=69$. The U-tiles are drawn in blue, the F-tiles in red and the R-tiles in green. The tiling was generated by using a generalization of the "domino-shuffling" exact sampling algorithm \cite{Janvresse2005,Doeraene} that allows for vanishing weights. The black curve is drawn using the parametric equation of subsection \ref{subsect_final_result}. The white dashed line delimits the deterministically frozen region.}
\label{fig_simulation_freezing_type1}
\end{figure}
\begin{figure}[h!]
\begin{center}
\includegraphics[scale=0.99]{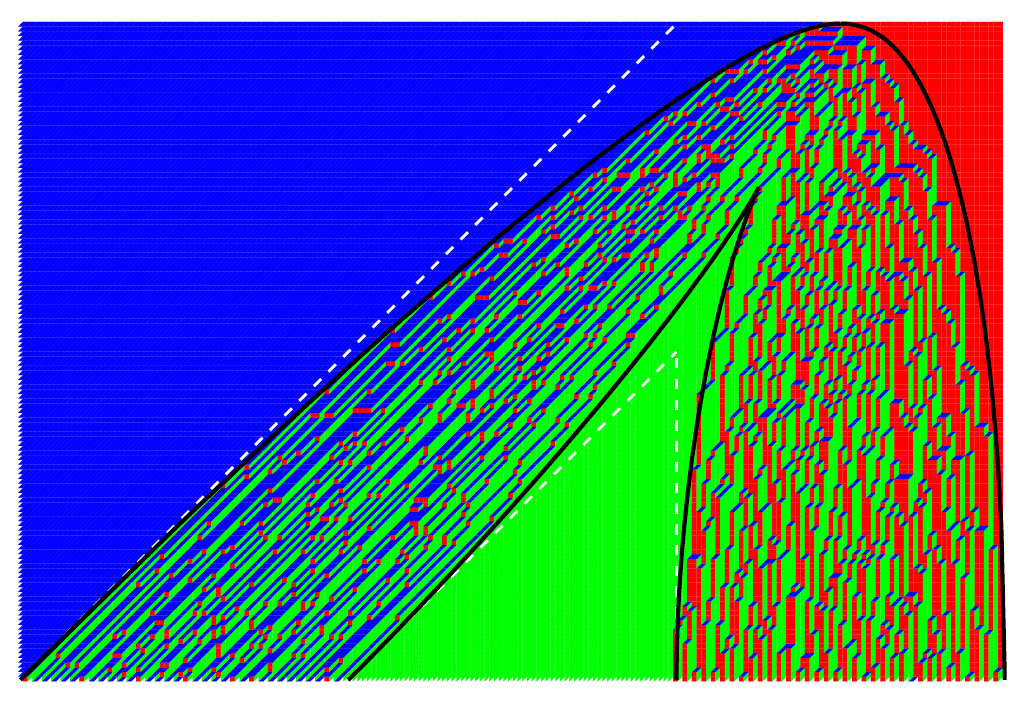}
\end{center}
\caption{Random tiling of a domain with a lower boundary composed of three intervals of equal length. The middle one presents a sawtooth pattern while the other two have  one defect every two sites. The number of paths (defects) is $n=140$. The tiling was generated by using a Markov chain algorithm. The black curve is obtained from the parametric equation of subsection \ref{subsect_final_result}. The white dashed lines delimit the deterministically frozen regions.}
\label{fig_simulation_freezing_type2}
\end{figure}

For non-generic distributions of $a_i$, more frozen regions, adjacent to the lower boundary of the rectangle, can appear. This is the case in particular when this distribution presents one (or several) so-called "freezing interval". These intervals can be of two types. Either it is a portion of the lower boundary with no starting points (see Figure \ref{fig_simulation_freezing_type1}) or a portion completely filled with starting points (see Figure \ref{fig_simulation_freezing_type2}). In the tiling picture, the interval is totally flat or has a sawtooth pattern. In either case, one observes a frozen region forming above the interval \cite{DiFrancesco2018ArbitraryStartingPoint}. Such a region is itself separated from the entropic bulk region by a new portion of arctic curve, whose shape depends on the chosen boundary condition. Its parametric representation has been given in \cite{DiFrancesco2018ArbitraryStartingPoint} where a proof by the tangent method has been provided in only one case, namely when the macroscopic interval has the sawtooth pattern and lies at one of the edges of the lower boundary. The rest of this note is devoted to giving a proof in the general case.

\section{Tangent method \label{sbsection_tangent_method}}

As pointed out above, each connected component of the frozen region is covered by tiles of one type only. The three types are treated separately, as they have a different path description in terms of which the frozen region is void of any path. A specific region in the entropic bulk that is adjacent to a frozen region is filled with paths, which, in the scaling limit, accumulate to form a portion of the arctic curve. The latter can thus be identified, after rescaling, with (any of) the outermost paths of the entropic region. For a frozen region formed above a freezing interval $I$, the outermost path starts from a point $O$ of the lower boundary of the domain, which coincides with the left or the right end of $I$.

Let us now move the starting point $O$ to another point $O'(\vec r)$, in such a way that the new random path has no other possibility than to enter the frozen region above the interval $I$ (the directedness and the non-intersection properties of the paths are crucial). The point $O'(\vec r)$ can be chosen to be inside the interval $I$, or in the frozen region itself, but the most convenient choice is to take it outside  the domain in which the model is defined (a rectangle in the present case). The way to force a path to start at $O'(\vec r)$ will be clear in the model at hand. This change transforms what was the outermost path of the entropic region into a new random path but otherwise should not affect the other outermost paths of the entropic zone, and therefore the arctic curve itself. The new random path, starting from $O'(\vec r)$ chosen outside the domain, will enter the domain through $I$, traverse the frozen region (actually being the only one to do so) and eventually hit the arctic curve.

The tangency assumption \cite{Colomo2016TangentMethod}, underlying the tangent method, states that in the scaling limit, the new path will almost surely be a straight line hitting the arctic curve tangentially, and from that point onward, will coincide with the arctic curve (as does the original outermost path). In models where the paths are subjected to no other interaction than the non-intersecting property, the tangency assumption has been recently proved in \cite{DGR19}. By varying the starting point $O'(\vec r)$, we obtain a family of lines which are tangent to the arctic curve, which can then be retrieved as the envelope of this family of straight lines. 

Given $O'(\vec r)$, chosen in the present case below the lower boundary of the domain, the straight line that eventually reaches the arctic curve is determined by the entry point of the path in the interval $I$. That point, say located  at a distance $\ell$ from $O$, may be explicitly computed, at finite volume, as the most likely entry point into the domain. Varying the starting points $O'(\vec r)$ yields a family of tangent lines parametrized by $\ell$, which, in the scaling limit, becomes continuous.

\section{Tangent method derivation for freezing boundaries}

A natural way to tackle the determination of the arctic curve would be to compute bulk expectation values to detect the frozen regions. It is however a much simpler task to use the tangent method \cite{Colomo2016TangentMethod}. The reason is that computing the most likely entry point only requires the knowledge of a boundary one-point function which is usually easier to evaluate. Moreover, for our particular model, this task can be greatly simplified by the Gelfand-Tsetlin formula \cite{Cohn1998shape}. This finite-size formula enumerates the tilings for our problem, even though it involves an arbitrary distribution of defects. The trick to fully exploit this formula is to extend the domain "from below".

\subsection{Freezing boundary of first kind: macroscopic flat interval \label{sbsection_freezing1}}

We first consider the case where the defect distribution contains a macroscopic flat interval, corresponding to an empty gap in the distribution of the starting points $O_i=(a_i,0)$ for the NILP of type 1 (see Figure \ref{fig_type_1_gap_extension}). In other words, there is an integer $q<n$ such that $a_{q+1}-a_q=m$, with $m$ proportional to $n$. We assume that the density to the left of $a_{q+1}$ and to the right of $a_q$ is strictly greater than 0. Above this flat interval two disjoint statistically frozen regions appear, respectively composed of F-tiles, and U-tiles (see Figure \ref{fig_simulation_freezing_type1}). We now study the corresponding F-portion and U-portion of the arctic curve separately, using the tangent method. The F-portion can be directly investigated using NILP of the first kind, while the U-type is more conveniently studied with the NILP of the second kind.

\subsubsection{F-portion}

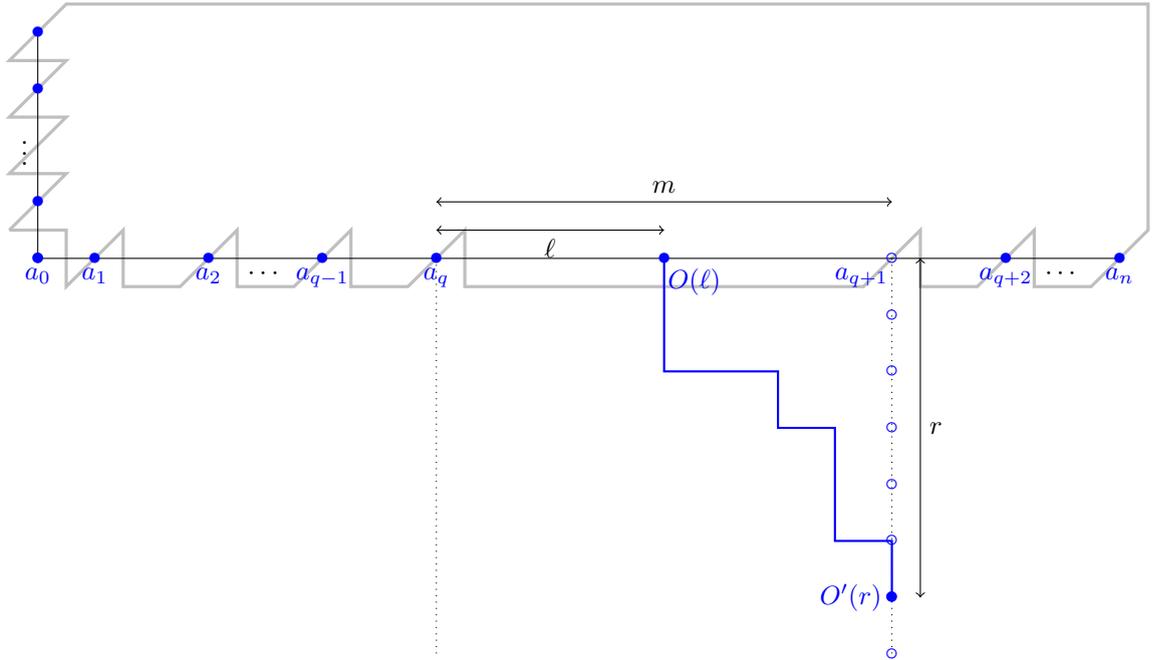
\begin{figure}[h!]
\begin{center}
\begin{tikzpicture}[scale=0.75]
\begin{scope}[xshift=-0.5cm,yshift=-0.5cm]
\draw[very thick, gray!50] (0,1)--++(1,0)--++(0,-1)--++(1,1)--++(0,-1)--++(1,0)--++(1,1)--++(0,-1)--++(0.5,0) node{}--++(0.5,0)--++(1,1)--++(0,-1)--++(1,0)--++(1,1)--++(0,-1)--++(7,0)--++(1,1)--++(0,-1)--++(1,0)--++(1,1)--++(0,-1) --++(0.5,0) node{}--++(0.5,0)--++(1,1)--++(0,1.5) node[below]{} --++(0,2.5)--++(-1.5,0) node{}--++(-14,0)node{}--++(-3.5,0)--++(-1,-1)--++(1,0)--++(-1,-1)--++(1,0)--++(-0.5,-0.5) node{}--++(-0.5,-0.5)--++(1,0)--++(-1,-1);
\end{scope}
\draw (0,0)--(0,4) node[midway,left]{$\vdots$};
\draw (0,0)--(19,0);
\draw [blue] (0,0) node{$\bullet$} node[below]{$a_0$};
\draw [blue] (1,0) node{$\bullet$} node[below]{$a_1$};
\draw [blue](3,0) node{$\bullet$} node[below]{$a_2$};
\draw (4,0) node[below]{$\cdots$};
\draw [blue](5,0) node{$\bullet$} node[below]{$a_{q-1}$};
\draw [blue](7,0) node{$\bullet$} node[below]{$a_q$};
\foreach \x in {0,1,3,4} {
\draw [blue](0,\x) node{$\bullet$};}
\draw[<->] (7,0.5)--(11,0.5) node[midway, below]{$\ell$};
\draw[<->] (7,1)--(15,1) node[midway, above]{$m$};
\draw [blue](15,0) node{$\circ$}node[below,xshift=-0.4cm]{$a_{q+1}$};
\draw [blue](17,0) node{$\bullet$} node[below]{$a_{q+2}$};
\draw (18,0) node[below]{$\cdots$};
\draw [blue](19,0) node{$\bullet$} node[below]{$a_{n}$};
\draw [dotted](15,0)--(15,-7);
\draw[dotted] (7,0)--(7,-7);
\foreach \x in {1,2,...,7} 
{\draw [blue] (15,-\x) node{$\circ$};}
\draw [thick,blue] (15, -6) node{$\bullet$} node[left]{$O'(r)$} --++(0,1)--++(-1,0)--++(0,1)--++(0,1)--++(-1,0)--++(0,1)--++(-1,0)--++(-1,0)--++(0,1)--++(0,1) node{$\bullet$} node[xshift=+0.4cm,below]{$O(\ell)$} ;
\draw[<->] (15.5,0)--(15.5,-6) node[midway,right]{$r$};
\end{tikzpicture}
\end{center}
\vspace{-0.5cm}
\caption{Extension of the domain by a semi-infinite strip to apply the tangent method using NILP of first type. The grey thick line represents the boundary of the original domain (without the extension) to ease the visualization of the starting points. The open dots ($\circ$) represent the possible starting positions of $O'(r)$ of the new path associated with $r=0,1, \cdots$. Note that for $r$ fixed, $\ell$ is random (at least at finite size).}
\label{fig_type_1_gap_extension}
\end{figure}

To apply the tangent method we extend the domain by a semi-infinite strip as indicated on Figure \ref{fig_type_1_gap_extension}. We move the point $O_{q+1}$ to  $O'(r)=(a_{q+1}, -r)$.
We consider the ratio of the partition function of this modified problem to the one of the original problem. It can be written as a sum over the entry point $O(\ell)=(a_q+\ell,0)$: 
\begin{equation}
\frac{Z(\{a_i\}_{i\neq q+1}, O'(r))}{Z(\{a_i\})}=\sum^m_{\ell=1} Y_{r,\ell} \; \frac{Z(\{a_i\}_{i\neq q+1}, O(\ell))}{Z(\{a_i\})} = \sum_{\ell=1}^m Y_{r,\ell} \; H_\ell,
\label{eq_ratio_partition_function_Ftype}
\end{equation}
where $Y_{r,\ell}$ counts the number of configurations of the path between $O'(r)$ and $O(\ell)$. We also introduced $Z(\{a_i\}_{i\neq q+1}, O(\ell))=Z(\{a'_i\})$ that counts the number of configurations in the original domain, but with a slightly modified situation: the starting points are $\{a'_i\}$ with $a'_i=a_i$ for $i \neq q+1$ and $a'_{q+1}=a_q+\ell$.  Following the tangent method recalled in section \ref{sbsection_tangent_method}, we need to find the most likely value of $\ell$. This can be achieved by doing a saddle point analysis of \eqref{eq_ratio_partition_function_Ftype}. 

To proceed, we first compute the quantities appearing on the right-hand side of \eqref{eq_ratio_partition_function_Ftype}. In the strip, the path has $m-\ell$ west-oriented elementary steps and $r$ north-oriented ones, where the last step leading to $O(\ell)$ has to be north-oriented. We therefore have
\begin{equation}
Y_{r,\ell}={m-\ell+r-1 \choose m-\ell}.
\end{equation}
The boundary one-point function $H_\ell$ can be easily determined by using the Gelfand-Tsetlin formula \cite{Cohn1998shape} (it was rederived in \cite{DiFrancesco2018ArbitraryStartingPoint} using the Lindström-Gessel-Viennot lemma and a LU decomposition). Computing $H_\ell$ is usually the most involved step in a tangent method derivation, but the Gelfand-Tsetlin formula makes it straightforward in the present case,
\begin{equation}
\begin{split}
H_\ell&=\frac{Z(\{a'_i\})}{Z(\{a_i\})}=\prod_{i=1}^n \prod_{s=0}^{i-1} \left(\frac{a'_i-a'_s}{a_i-a_s} \right)\\
&=\exp\left\{\sum_{i=0}^q \log \left(1+\frac{m-\ell}{a_i-a_{q+1}} \right)+\sum_{i=q+2}^n \log \left(1+\frac{m-\ell}{a_i-a_{q+1}} \right)\right\}.
\end{split}
\label{eq_boundary_one_point_fct}
\end{equation}

We now consider the rescaled domain where we divide all lengths by $n$. We introduce the rescaled variables
\begin{equation}
\ell=n\xi, \quad q=nu_1, \quad m=n\delta, \quad i=nu, \quad r=nz, \quad a_{i} = n \alpha\left(\frac{i}{n}\right),
\end{equation}
where the function $\alpha(u)$ characterises the defect distribution. Since $a_{q+1}=a_q+m$, in the $n \to \infty$ limit, the function $\alpha$ will have a discontinuity at $u_1$ where it jumps by an amount $\delta$. Thus we set $\alpha(u_1) = \lim_{u \to u_1^{-}} \alpha(u)$ while $\lim_{u \to u_1^{+}} \alpha(u) = \alpha(u_1)+\delta$. In the limit $n \to \infty$, we have $Y_{r,\ell} \sim e^{nS_0}$ and $H_\ell \sim e^{nS_1}$ with 
\begin{equation}
S_0=(\delta-\xi+z) \log(\delta-\xi+z) - (\delta-\xi)\log(\delta-\xi)-z\log z
\end{equation}
and
\begin{equation}
\begin{split}
S_1 &=  \int_0^{1} \dif u \; \log \left( 1+\frac{\delta-\xi}{\alpha(u)-(\alpha(u_1)+\delta)}\right).
\end{split}
\end{equation}
We perform the saddle point analysis on
\begin{equation}
\sum_{\ell=1}^m Y_{r,\ell} \; H_\ell \sim n \int_0^{\delta} \dif \xi \; e^{n[S_0+S_1]}. 
\label{eq_steepest_1}
\end{equation}
Let us denote by $\xi^*$ the solution to the saddle point equation $\frac{\pa}{\pa \xi} (S_0+S_1)=0$. We could choose to parametrize the family of tangent lines by $z$, viewing $\xi^*=\xi^*(z)$ as a function of $z$, but it turns out to be more convenient to use the intercept $t=\alpha(u_1)+\xi^*$ of the tangent line with the $X-$axis, thereby making $z=z(t)$ a function of $t$. The arctic curve will then be parametrized by $t$ as well. Defining
\begin{equation}
x(t) \equiv \exp\left\{ -\int_0^1 \dif u \; \frac{1}{t-\alpha(u)} \right\},
\label{eq_defxt}
\end{equation}
the saddle point condition yields
\begin{equation}
\frac{\alpha(u_1) + \delta - t}{\alpha(u_1) + \delta - t + z(t)} = x(t).
\label{eq_stationarity_F_type}
\end{equation}
By definition the parameter $t$ must be contained in the interval $[\alpha(u_1),\alpha(u_1)+\delta] = [\alpha(u_1^-),\alpha(u_1^+)]$, but not all values in this interval will correspond to real positive values of $z$.


Let us write
\begin{equation}
I(t) \equiv - \log{x(t)} = \int_0^{u^-_1} \dif u \; \frac{1}{t-\alpha(u)} - \int_{u_1^+}^1 \dif u \; \frac{1}{\alpha(u)-t}.
\end{equation}
When $t$ is in the interior of $[\alpha(u_1),\alpha(u_1)+\delta]$, it is not in the image of the function $\alpha(u)$, so that $I(t)$ is a well-defined, continuous and decreasing function. Moreover $I(t)$ diverges to $+\infty$ when $t$ goes to the lower bound $\alpha(u_1^-)$, and to $-\infty$ when $t$ goes to the upper bound $\alpha(u_1^+)$.  Therefore, on the closed interval, $I(t)$ takes all values from $-\infty$ to $+\infty$, and $x(t)$ itself is a continuous increasing function, taking all positive values from 0 to $+\infty$. However from (\ref{eq_stationarity_F_type}), $z$ being real and positive implies that $x(t)$ is between 0 and 1. Therefore the proper range of $t$ to consider is the interval $[\alpha(u_1),t_1]$ where $t_1 < \alpha(u_1)+\delta$ is defined by $x(t_1)=1$. We then have $z(\alpha(u_1))=+\infty$ and $z(t_1)=0$, corresponding respectively to a vertical tangent and a horizontal tangent. We will indeed check that $t_1$ corresponds to the transition point from the F-frozen region to the U-frozen one. 

Imposing in the rescaled domain that the tangent line associated with the parameter $t$ passes through the points $(t,0)$ and $(\alpha(u_1)+\delta, -z(t))$ leads to the equation 
\begin{equation}
x(t)Y+(1-x(t))(X-t)=0, \quad t \in [\alpha(u_1),t_1],
\label{eq_family_lines}
\end{equation}
which is exactly the equation (3.9) of \cite{DiFrancesco2018ArbitraryStartingPoint}.

\subsubsection{U-portion}

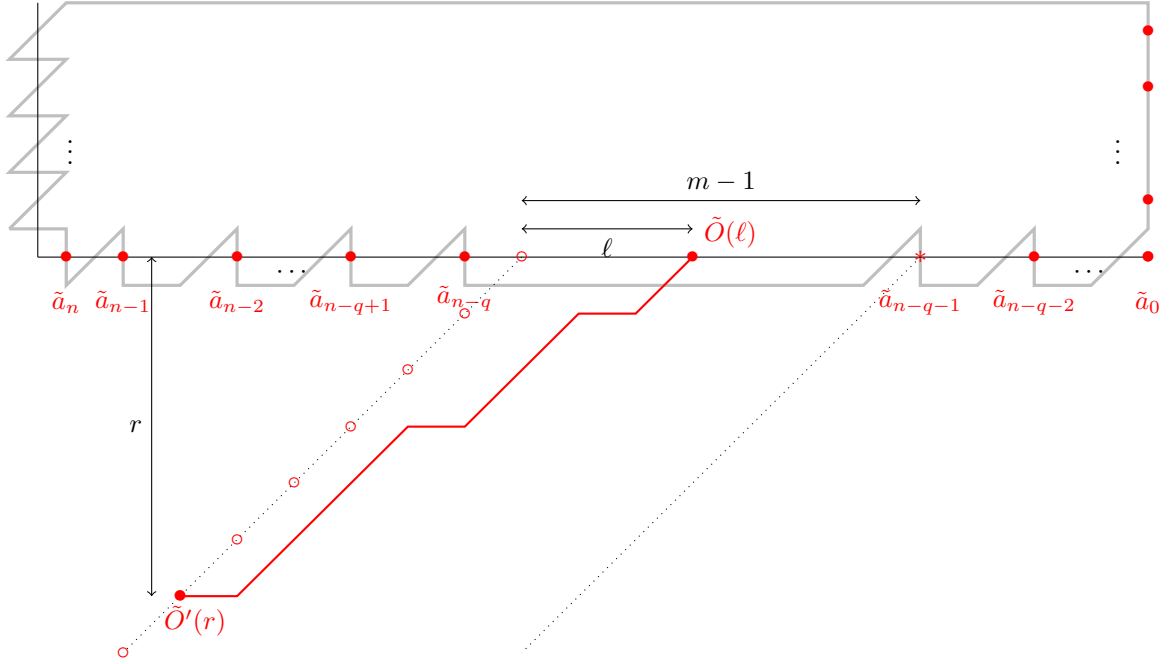
\begin{figure}[t]
\begin{center}
\begin{tikzpicture}[scale=0.75]
\begin{scope}[xshift=0cm,yshift=-0.5cm]
\draw[very thick, gray!50] (0,1)--++(1,0)--++(0,-1)--++(1,1)--++(0,-1)--++(1,0)--++(1,1)--++(0,-1)--++(0.5,0) node{}--++(0.5,0)--++(1,1)--++(0,-1)--++(1,0)--++(1,1)--++(0,-1)--++(7,0)--++(1,1)--++(0,-1)--++(1,0)--++(1,1)--++(0,-1) --++(0.5,0) node{}--++(0.5,0)--++(1,1)--++(0,1.5) node[below]{} --++(0,2.5)--++(-1.5,0) node{}--++(-14,0)node{}--++(-3.5,0)--++(-1,-1)--++(1,0)--++(-1,-1)--++(1,0)--++(-0.5,-0.5) node{}--++(-0.5,-0.5)--++(1,0)--++(-1,-1);
\end{scope}
\draw (20,2) node[xshift=-0.4cm]{$\vdots$};
\draw (0,2) node[xshift=0.8cm]{$\vdots$};
\draw (0.5,0)--(20,0);
\draw (0.5,0)--(0.5,4.5);
\begin{scope}[xshift=1cm]
\draw [red](0,0) node{$\bullet$} node[yshift=-0.6cm]{$\tilde{a}_n$};
\draw [red](1,0) node{$\bullet$} node[yshift=-0.6cm]{$\tilde{a}_{n-1}$};
\draw [red](3,0) node{$\bullet$} node[yshift=-0.6cm]{$\tilde{a}_{n-2}$};
\draw (4,0) node[below]{$\cdots$};
\draw [red](5,0) node{$\bullet$} node[yshift=-0.6cm]{$\tilde{a}_{n-q+1}$};
\draw [red](7,0) node{$\bullet$} node[yshift=-0.55cm]{$\tilde{a}_{n-q}$};
\foreach \x in {0,1,3,4} {
\draw [red](19,\x) node{$\bullet$};}
\draw[<->] (8,0.5)--(11,0.5) node[midway, below]{$\ell$};
\draw[<->] (8,1)--(15,1) node[midway, above]{$m-1$};
\draw [red] (15,0) node{$\ast$}node[yshift=-0.6cm]{$\tilde{a}_{n-q-1}$};
\draw [red] (17,0) node{$\bullet$} node[yshift=-0.6cm]{$\tilde{a}_{n-q-2}$};
\draw (18,0) node[below]{$\cdots$};
\draw [red] (19,0) node{$\bullet$} node[yshift=-0.6cm]{$\tilde{a}_0$};
\end{scope}
\draw [dotted](16,0)--(16-7,-7);
\draw[dotted] (9,0)--(9-7,-7);
\foreach \x in {0,1,2,...,7} 
{\draw[red] (9-\x,-\x) node{$\circ$};}
\draw[thick,red] (9-6, -6) node{$\bullet$} node[below,xshift=0.2cm]{$\tilde{O}'(r)$} --++(1,0)--++(1,1)--++(1,1)--++(1,1)--++(1,0)--++(1,1)--++(1,1)--++(1,0)--++(1,1)node{$\bullet$} node[xshift=+0.5cm,above]{$\tilde{O}(\ell)$} ;
\draw[<->] (9-6-0.5,0)--(9-6-0.5,-6) node[midway,left]{$r$};
\end{tikzpicture}
\end{center}
\vspace{-0.5cm}
\caption{Extension of the domain with a semi-infinite strip to apply the tangent method using NILP of second kind. The starting points $\tilde{O}_i=(\tilde{a}_i,0)$ and the ending points are different for these paths, as represented (see \cite{DiFrancesco2018ArbitraryStartingPoint} for more details). The star ($*$) represents the position of the starting point that we move to $\tilde{O}'(r)$.}
\label{fig_type_2_gap_extension}
\end{figure}

We now consider the U-portion. In what follows, we will use the same names ($r$, $Y_{r,\ell}$, $z$ and so on) for quantities that are conceptually identical but which take different values. To use the tangent method, we consider an equivalent description, in terms of NILP of second kind (see Figure \ref{fig_type_2_gap_extension}).  We extend the domain in a similar fashion as for the F-case by gluing a semi-infinite tilted strip and by moving $\tilde{O}_{n-q-1}$ to $\tilde{O}'(r)=(\tilde{a}_{n-q}+1-r,-r)$. The computation is quite similar to the F-type case. The boundary one-point function can be computed by using NILP of first kind and is again given by \eqref{eq_boundary_one_point_fct}, while the number of configurations of the path (of second kind) between $\tilde{O}'(r)$ and $\tilde{O}(\ell)=(\tilde{a}_{n-q}+1+\ell,0)$ is 
\begin{equation}
Y_{r,\ell}={r+\ell-1 \choose \ell},
\end{equation}
where we imposed that the last step is north-east-oriented. Using the same notation for the rescaled variables, we find that asymptotically $Y_{r,\ell}\sim e^{nS_0}$ with 
\begin{equation}
S_0=(z+\xi)\log(z+\xi)-\xi\log\xi-z\log z.
\end{equation}

From the saddle point analysis, we find that
\begin{equation}
\frac{t-\alpha(u_1)+z(t)}{t-\alpha(u_1)} = x(t),
\label{eq_saddlepoint_Uportion}
\end{equation}
for the same continuous increasing function $x(t)$ given in (\ref{eq_defxt}). From \eqref{eq_saddlepoint_Uportion}, $z(t)$ being real and positive implies that $x(t)$ takes values from $1$ to $\infty$. Therefore the proper range of $t$ to consider is $[t_1,\alpha(u_1)+\delta]$ with $x(t_1)=1$. 

By imposing that the tangent line of parameter $t$ passes through $(t,0)$ and $(\alpha(u_1)-z(t),-z(t))$, we find
\begin{equation}
x(t)Y+(1-x(t))(X-t)=0,\quad t\in[t_1, \alpha(u_1)+\delta],
\end{equation}
which is the same expression as in \eqref{eq_family_lines}. The proof is hence complete for freezing boundaries of first kind, since the envelope of this family of lines will have the same parametric representation as in \cite{DiFrancesco2018ArbitraryStartingPoint}, but with $t \in [\alpha(u_1),\alpha(u_1)+\delta]$.

\subsection{Freezing boundary of second kind: macroscopic sawtooth pattern \label{sect_freezing_second_kind}}

We now turn to the tiling problem when a macroscopic portion of the lower boundary has a sawtooth pattern. At the level of the description in terms of NILP of first kind, it corresponds to an interval of indices $I=\{q,q+1, \cdots , q+m-1\}$ for which the starting points $O_i=(a_i,0)$ are tightly arranged: $a_{i+1}-a_i=1$ for $i \in I$. We assume that the density to the left of $a_q$ and to the right of $a_{q+m}$ is strictly less than 1 over a macroscopic distance. We further suppose that there is no starting point directly adjacent to this region, namely, $a_q-a_{q-1}$ and $a_{q+m}-a_{q+m+1}$ are both greater than $1$. If the freezing interval is not directly adjacent to the left or right corner of the domain, this condition is simply a convenient way to define the interval. We will discuss the case when the freezing interval is on the left or right border. Again, by macroscopic portion, we mean that $m$ scales like $n$. In such a situation, there is a deterministically frozen triangular region arising in a similar fashion as the one associated with the ending points discussed in the second section, and around it, also a region that is statistically frozen. The whole frozen region is covered by an arrangement of R-tiles. We now study the associated arctic curve using the tangent method with the third kind of NILP for which the region of interest is empty. 

\begin{figure}[h!]
\begin{center}
\begin{tikzpicture}[scale=0.75]
\begin{scope}[xshift=-0.5cm,yshift=-0.5cm]
\draw[very thick, gray!50] (0,1)--++(1,0)--++(0,-1)--(1,0)--++(1,0)--++(1,1)--++(0,-1)--++(2,0)--++(1,1)--++(0,-1)--++(0.5,0) node{}--++(0.5,0)--++(1,1)--++(0,-1)--++(1,0)--++(1,1)--++(0,-1)--++(1,1)--++(0,-1)--++(1,1)--++(0,-1)--++(1,1)--++(0,-1)--++(1,1)--++(0,-1)--++(1,1)--++(0,-1)--++(1,1)--++(0,-1)--++(1,0)--++(0.5,0.5)node{}--++(0.5,0.5)--++(0,-1) --++(1,0)--++(1,1) --++(0,3) node[below]{} --++(0,3)--++(-1.5,0) node{}--++(-14,0)node{}--++(-3.5,0)--++(-1,-1)--++(1,0)--++(-1,-1)--++(1,0)--++(-0.5,-0.5) node{}--++(-0.5,-0.5)--++(1,0)--++(-1,-1)--++(1,0)--++(-1,-1)--++(1,0)--++(-1,-1)--++(1,0);
\end{scope}
\draw (0,0)--(0,6.5);
\draw (0,0)--(19,0);
\draw (20-0.5,3) node[xshift=-0.4cm]{$\vdots$};
\draw (0-0.5,3) node[xshift=0.8cm]{$\vdots$};
\draw [green] (1,-0.5) node{$\bullet$} node[yshift=0.4cm]{$|$} node[yshift=0.8cm]{$b_1$};
\draw [green](3,-0.5) node{$\bullet$} node[yshift=0.4cm]{$|$} node[yshift=0.8cm]{$b_2$};
\draw [green](4,-0.5) node{$\bullet$} node[yshift=0.4cm]{$|$} node[yshift=0.8cm]{$b_3$};
\draw (6,-0.5) node[below]{$\cdots$};
\draw [green](8,-0.5) node{$\bullet$} node[yshift=0.4cm]{$|$} node[yshift=0.8cm]{$b_{p-1}$};
\draw [green] (16,-0.5) node{$\circ$} node[yshift=0.4cm]{$|$} node[yshift=0.8cm]{$b_p$};
\draw [dotted](16,-0.5)--++(-7,-7);
\draw [dotted](16-7,-0.5)--++(-7,-7);
\foreach \x in {1,2,...,7} 
{\draw [green] (16-\x,-0.5-\x) node{$\circ$};}
\draw[thick,green] (16-6, -0.5-6) node{$\bullet$} node[right]{$\hat{O}'(r)$} --++(0,1)--++(1,1)--++(0,1)--++(0,1)--++(1,1)--++(0,1) node{$\bullet$} node[right,yshift=-0.3cm]{$\hat{O}(\ell)$} ;
\draw[<->] (16,-0.5)--(16,-0.5-6) node[midway,right]{$r$};
\draw[<->] (8+1,0.6)--(12,0.6) node[midway, above]{$\ell$};
\draw[<->] (8+1,1.2)--(16,1.2) node[midway, above]{$m$};
\draw (17,-0.5) node[below]{$\cdots$};
\draw [green] (18,-0.5) node{$\bullet$} node[yshift=0.4cm]{$|$} node[yshift=0.8cm]{$b_{\tilde{m}}$};
\foreach \x in {1,2,3,5,6,8} \draw [green] (10+\x,6+0.5) node{$\bullet$};
\draw (10+4,6+0.5) node[above]{$\cdots$};
\draw (10+7,6+0.5) node[above]{$\cdots$};
\end{tikzpicture}
\end{center}
\vspace{-0.5cm}
\caption{Extension of the domain to apply the tangent method, using NILP of third kind. The original starting points are $\hat{O}_i=(b_i,-1/2)$ for $i=1, 2, \cdots, \tilde{m}$ with $\tilde{m}=a_n-n$ and the ending points are at $(n+i, n+1/2)$. The parameter $p$ gives the position of the freezing interval in this NILP description, but we do not need to know its precise relation to $q$ (see again \cite{DiFrancesco2018ArbitraryStartingPoint} for more details).}
\label{fig_type_3_tighly_extension}
\end{figure}
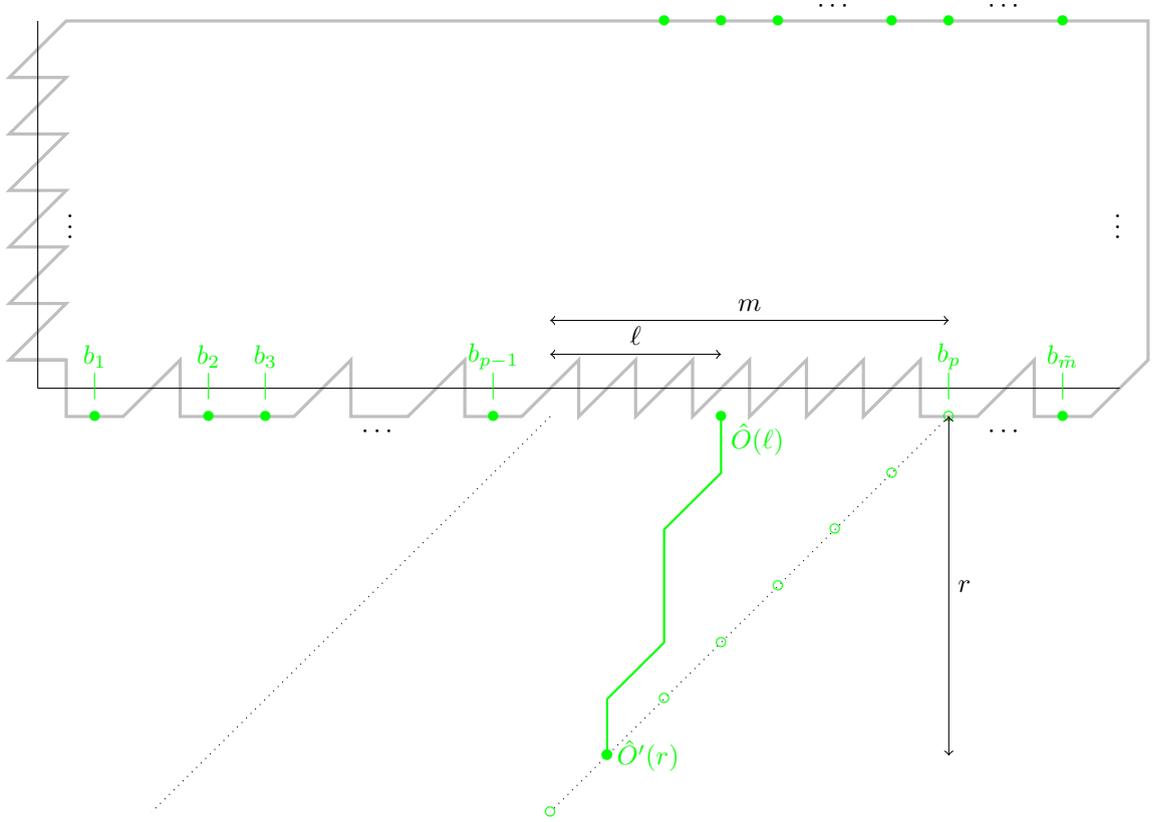
We extend the domain with a semi-infinite tilted strip (see Figure \ref{fig_type_3_tighly_extension}). We move $\hat{O}_p$ to $\hat{O}'(r)=(b_p-r, -r-1/2)$. The same extension can be done if the freezing interval touches the left edge of the bottom boundary ($q=0$). If it touches the right edge ($q+m=n$), we take for $\hat{O}_p$ the starting point of an ad-hoc trivial vertical path of third type added on the right of the domain. As before, we consider the quantity 
\begin{equation}
\frac{Z(\{b_i\}_{i\neq p}, \hat{O}'(r))}{Z(\{b_i\})}=\sum_{\ell=0}^m Y_{r,\ell} \; \frac{Z(\{b_i\}_{i\neq p}, \hat{O}(\ell))}{Z(\{b_i\})} = \sum_{\ell=0}^m Y_{r,\ell} \; H_\ell,
\end{equation}
where $Y_{r,\ell}$ counts the number of configurations of the path of third type between $\hat{O}'(r)$ and $\hat{O}(\ell)=(b_{p-1}+1+\ell, -1/2)$ and $Z(\{b_i\}_{i\neq p}, \hat{O}(\ell))$ counts the number of configurations in the original domain, but with a slightly modified bottom boundary.  Since the path cannot go to the left, we have $Y_{r,\ell}=0$ when $\ell < m-r$. When $\ell \ge m-r$ we find that

\begin{equation}
Y_{r,\ell}= {r \choose r+\ell-m}.
\end{equation}
Instead of working with the third kind of NILP to evaluate $H_\ell$, we can also use the first kind of NILP to do so,
\begin{equation}
H_\ell = \frac{Z(\{a'_i\})}{Z(\{a_i\})}, 
\end{equation}
with the following new starting points of the NILP of first kind,
\begin{equation}
\begin{split}
&a'_i=a_i, \quad  \quad \quad \text{for} \quad i \not\in \{q+\ell+1, q+\ell+2, \cdots, q+m\},\\
&a'_{i}=a_i+1, \quad \text{for} \quad i \in \{q+\ell+1, q+\ell+2, \cdots, q+m\},
\end{split}
\end{equation}
%
which simply means that all the points from $a_{q+\ell+1}$ to $a_{q+m}$ are shifted by one unit to the right. Using again the Gelfand-Tsetlin formula, we find that
\begin{equation}
\begin{split}
H_\ell&=\prod_{i=1}^n \prod_{s=0}^{i-1} \left(\frac{a'_i-a'_s}{a_i-a_s} \right)=\prod_{i=q+\ell+1}^{q+m} \; \prod_{s=0}^{q+\ell} \left(\frac{a_i-a_s+1}{a_i-a_s}\right)  \prod_{i=q+m+1}^n \; \prod_{s=q+\ell+1}^{q+m} \left( \frac{a_i-a_s-1}{a_i-a_s}\right)\\
&=\exp\left\{ \sum_{i=q+\ell+1}^{q+m} \; \sum_{s=0}^{q+\ell} \log \left(\frac{a_i-a_s+1}{a_i-a_s}\right) + \sum_{i=q+m+1}^n \; \sum_{s=q+\ell+1}^{q+m} \log\left( \frac{a_i-a_s-1}{a_i-a_s}\right)\right\},
\end{split}
\label{eq_Hl_2nd_kind}
\end{equation}
where we expressed this quantity in terms of the original $(a_i)_i$. We used the fact that the factor $a_i'-a_s'$ in the numerator can only differ from the factor $a_i-a_s$ in the denominator by an amount equal to $+1$ , $-1$ or $0$. 

We introduce the rescaled variables as follows, 
\begin{equation}
 \ell=n\xi , \quad r=nz, \quad q=n u_1, \quad m=n u_2, \quad i=nu, \quad s=nv, \quad a_i = n \alpha\left(\frac{i}{n} \right),
\end{equation}
with the condition
\begin{equation}
\xi \ge u_2 - z.
\label{cond}
\end{equation}
In the limit $n \to \infty$ the function $\alpha(u)$ will be such that $\alpha'(u)=1$ for $u\in[u_1,u_1+u_2]$, so that $\alpha(u_1+\xi) = \alpha(u_1)+\xi$ for any $\xi$ in $[0,u_2]$. We also note that the condition $a_q-a_{q-1} > 1$ implies that the left derivative of $\alpha$ at $u_1$ is strictly greater than 1, and likewise for the right derivative at $u_1+u_2$,
\begin{equation}
\lim_{u \to u_1^-}\alpha'(u) > 1, \quad \lim_{u \to u_1^+}\alpha'(u) = 1, \qquad 
\lim_{u \to (u_1+u_2)^-} \alpha'(u) = 1, \quad \lim_{u \to (u_1+u_2)^+} \alpha'(u) > 1.
\end{equation}

In the large $n$ limit, we have $Y_{r,\ell} \sim e^{n S_0}$ and $H_\ell \sim e^{n S_1}$ with 
\begin{equation}
\begin{split}
S_0 &=  z \log z - (z+\xi-u_2)\log(z+\xi-u_2) -(u_2-\xi) \log( u_2-\xi),\\
S_1&=  \int_{u_1 + \xi + \epsilon}^{u_1+u_2} \dif u \; \int_{0}^{u_1 + \xi} \dif v \; \frac{1}{\alpha(u)-\alpha(v)} - \int_{u_1 +u_2}^{1} \dif u \; \int_{u_1 + \xi }^{u_1+u_2-\epsilon} \dif v \; \frac{1}{\alpha(u)-\alpha(v)},
\end{split}
\label{eq_S1_2nd_kind}
\end{equation}
where $\epsilon=1/n$. The $\epsilon$ regularization can be read from the values taken by $i$ and $s$ in \eqref{eq_Hl_2nd_kind}. A careful analysis of the asymptotics of \eqref{eq_Hl_2nd_kind} requires to treat separately the terms in which $i$ is close to $s$. This can be done by using the property $a_s=a_q+(s-q)$ for $s \in I$ and by assuming that the density of $a_i$'s is constant (and strictly less than 1) on a macroscopic portion to the right of $a_{q+m}$. This computation exactly leads to \eqref{eq_S1_2nd_kind}. We apply the steepest descent method on 
\begin{equation}
\sum_{\ell=m-r}^m Y_{r,\ell} \; H_\ell \sim n \int_{u_2-z} ^{u_2} \dif \xi  \; e^{n(S_0+S_1)}.
\label{eq_steepest_secondfreezing}
\end{equation}
The derivatives of the actions $S_0$ and $S_1$ with respect to $\xi$ read
\begin{equation}
\begin{split}
\frac{\pa S_0}{\pa \xi} &= \log \left( \frac{u_2-\xi }{z+\xi-u_2}\right),\\
\frac{\pa S_1}{\pa \xi} &= \int_0^{u_1+\xi-\epsilon} \dif u \;  \frac{1}{\alpha(u)-\alpha(u_1+\xi)} + \int_{u_1+\xi+\epsilon}^1 \dif u \;  \frac{1}{\alpha(u)-\alpha(u_1+\xi)}\\
&= \text{p.v.} \int_0^1 \dif u \;\frac{1}{\alpha(u)-\alpha(u_1+\xi)}.  
\end{split}
\end{equation}
The stationarity condition gives  $\xi^*$, corresponding to the entry point of the tangent:
\begin{equation}
\frac{u_2-\xi^*}{z+\xi^*-u_2} \exp\left\{- \text{p.v.} \int_0^1 \dif u \;\frac{1}{\alpha(u_1+\xi^*)-\alpha(u)}\right\}=1.
\end{equation}
Let us again introduce $t=\alpha(u_1 +\xi^*) = \alpha(u_1)+\xi^*$ the intercept of the tangent line with the $X-$axis. We also define 
\begin{equation}
x(t)=- \exp\left\{- \text{p.v.} \int_0^1 \dif u \;\frac{1}{t-\alpha(u)}\right\},
\label{eq_absolute_value_xt}
\end{equation}
where we introduce the minus sign in the definition of $x(t)$ for this range of $t$ in order to match the parametric form of the arctic curve given in \cite{DiFrancesco2018ArbitraryStartingPoint}; it therefore implies $x(t)\le 0$. The above condition then becomes
\begin{equation}
z(t) = \big({\alpha(u_1+u_2)-t}\big) \big(1 - x(t)\big).
\label{eq_z1}
\end{equation}
The negativity of $x(t)$ ensures that the condition (\ref{cond}) is satisfied.

By definition $t$ must be in $[\alpha(u_1), \alpha(u_1+u_2)]$, but the whole interval may not be covered as we move $z$ from 0 to $+\infty$. Let us split the integral appearing in $x(t)$ as follows
\begin{equation}
I(t):=\int_0^1 \dif u \frac{1}{t-\alpha(u)} = \int_0^{u_1} \frac{\dif u}{t-\alpha(u)} + \int_{u_1}^{u_1+u_2} \frac{\dif u}{t-\alpha(u)} + \int_{u_1+u_2}^1 \frac{\dif u}{t-\alpha(u)},
\label{split}
\end{equation}
with the prescription that the singularity of the integrand at $u=\alpha^{-1}(t)$ must be approached symmetrically from the left and from the right.

When $t$ is in the open interval $]\alpha(u_1),\alpha(u_1+u_2)[$, the singularity is in the second integral, for which the prescription on how to approach the singularity returns the principal value of the integral. Using the linearity of $\alpha$ in the interval $[u_1,u_1+u_2]$, we find
\begin{equation}
\text{p.v.} \int_{u_1}^{u_1+u_2} \frac{\dif u}{t-\alpha(u)} =
\text{p.v.} \int_{u_1}^{u_1+u_2} \frac{\dif u}{t-(\alpha(u_1)+(u-u_1))}
=\log \left(\frac{t-\alpha(u_1)}{\alpha(u_1+u_2)-t} \right),
\end{equation}
which takes all real values when $t$ varies over the open interval. The other two integrals are finite for $t$ in the open interval, but diverge as $t$ approaches the lower and upper bounds of the interval.

When $t$ gets close to the lower bound $\alpha(u_1)$, the integrands of the first two integrals in (\ref{split}) are singular, respectively at $u_1^-$ and $u_1^+$. Expanding around these two points yields 
\begin{equation}
\frac1{\alpha(u_1)-\alpha(u)} \sim \frac1{\alpha'(u_1^\pm)} \frac1{u-u_1}, \qquad u \sim u_1^\pm.
\end{equation}
Applying the prescription to stay at a distance $\epsilon$ from the singularity, we see that the first two integrals both contain a logarithmically diverging term, equal to $-\log {\epsilon}/\alpha'(u_1^-)$ for the first integral, $\log {\epsilon}/\alpha'(u_1^+)$ for the second one. As $\alpha'(u_1^-)>1$ and $\alpha'(u_1^+)=1$, the two divergences do not cancel out. As $\epsilon$ goes to 0, their sum diverges to $\lim_{\epsilon \to 0} (1-1/{\alpha'(u_1^-)}) \log\epsilon = -\infty$. As a consequence, $I(\alpha(u_1))$ and $x(\alpha(u_1))$ both diverge to $-\infty$. A similar analysis for the upper bound $t \to \alpha(u_1+u_2)$ of the interval shows that $I(t)$ diverges to $+\infty$, and $x(t)$ tends to 0. By continuity the function $I(t)$ takes all values from $-\infty$ to $+\infty$ when $t$ is varied over the closed interval $[\alpha(u_1), \alpha(u_1+u_2)]$, and correspondingly $x(t)$ takes all values from $-\infty$ to 0.

It follows that $z(t)$ takes all values from $+\infty$ to 0 when $t$ runs over $[\alpha(u_1), \alpha(u_1+u_2)]$, equivalently the intercept $\xi^*$ covers the full interval $[0,u_2]$. In the rescaled domain, the equation of the tangent line of parameter $t$ passing through the points $(0,t)$ and $(\alpha(u_1)+u_2-z(t),-z(t))$ is then 
\begin{equation}
x(t)Y + \big(1-x(t)\big) (X-t) = 0, \quad t \in [\alpha(u_1),\alpha(u_1+u_2)].
\end{equation}
From the limits given above, the tangent has a slope 1 for $t=\alpha(u_1)$, and is vertical for $t=\alpha(u_1+u_2)$ (see Figure \ref{fig_simulation_freezing_type2}).

\subsection{Final results \label{subsect_final_result}}

In all cases, we conclude that
\begin{equation}
x(t)Y+(1-x(t))(X-t)=0,\quad \quad  t\in I_{\text{freezing}},
\end{equation}
for $I_\text{freezing}$ an interval of the real line where we have a freezing boundary of one of the two kinds discussed above. Retrieving the envelope of a family of tangent lines precisely corresponds to an inverse Legendre transform that can indeed be used to find the equation $X(Y)$ for different portions of the arctic curve \cite{DiFrancesco2018ArbitraryStartingPoint}. We can also give a parametric form of the curve
\begin{equation}
\left\{ 
\begin{split}
&x(t)Y+(1-x(t))(X-t)=0,\\
&x'(t)Y-x'(t)(X-t)-(1-x(t))=0.
\end{split}
\right.
\end{equation}
Solving the system above leads to the following equation for the arctic curve
\begin{equation}
\begin{split}
X(t) &= t-\frac{x(t)(1-x(t))}{x'(t)},\\
Y(t) &= \frac{(1-x(t))^2}{x'(t)}.
\end{split} \quad \quad \quad t\in I_{\text{freezing}}
\label{eq_prediction}
\end{equation}
The portions of the arctic curve that we have not analysed so far were derived in \cite{DiFrancesco2018ArbitraryStartingPoint} by extending the domain above the upper edge, whereas the derivation for general freezing boundaries requires an extension below the lower edge. Incidentally, let us mention that we can also access these remaining portions of arctic curves (with a range $t \in [\alpha(1),\infty[$ and $t\in]-\infty,0]$ )  by extending the domain from below, in the following way. For $t \in [\alpha(1),\infty[$ we consider NILP of first kind and attach on the right of the original domain a semi-infinite horizontal strip whose upper horizontal edge is aligned with that of the original domain and of width $(1+a)n$ with $a>0$ arbitrary. We then move $O_n$ to the lower horizontal edge of this extension. For $t\in]-\infty,0]$, we consider NILP of second kind and use a similar semi-infinite horizontal strip, this time attached on the left, also making $\tilde{O}_n$ start from the lower edge of the strip. In both cases, we recover $x(t)Y+(1-x(t))(X-t)=0$ by using the same kind of arguments as above.

\section{Conclusion}

We have considered the arctic curve for the tiling model defined in \cite{DiFrancesco2018ArbitraryStartingPoint} by applying the tangent method with an extension of the domain from below. This extension enables the determination of all the portions of arctic curve (including the ones not covered in \cite{DiFrancesco2018ArbitraryStartingPoint}) in a systematic way, by using the Gelfand-Tsetlin formula. The resulting parametric equation of the arctic curve has been compared with random tilings obtained from numerical simulations, for some particular choices of the defect distribution.

\paragraph{Acknowledgments} We are grateful to P. Di Francesco and E. Guitter for their encouragement to publish this proof. We are indebted to Antoine Doeraene who kindly accepted to implement the generalized shuffling algorithm for this particular problem (other random tilings can be generated on the website \cite{Doeraene}).  We also thank Gilles Parez and anonymous referees for their careful reading of the manuscript. B.D. acknowledges the financial support of the Fonds Sp\'eciaux de Recherche (FSR) of the Universit\'e catholique de Louvain. P.R. is a Senior Research Associate of FRS-FNRS (Belgian Fund for Scientific Research). This work was supported by the Fonds de la Recherche Scientifique-FNRS under the EOS-contract O013018F.

\end{document}